\documentclass[comment, prl,twocolumn,nofootinbib, preprintnumbers, superscriptaddress]{revtex4}
\usepackage{amsmath,amssymb,bm,slashed,braket}
\usepackage{graphicx}
\usepackage{epstopdf}
\usepackage{float}
\usepackage{xcolor}
\usepackage{multirow}
\usepackage{dsfont}
\usepackage[colorlinks=true,
            linkcolor=blue,
            urlcolor=blue,
            citecolor=purple,          
            bookmarks=true,
            bookmarksnumbered=true,
            breaklinks=true,
            pdfpagemode=Fullscreen,
            pdfstartview=FitBH]{hyperref}
\usepackage[normalem]{ulem}
\usepackage{orcidlink}
\allowdisplaybreaks[4]
\usepackage[capitalise]{cleveref}

\begin{document}

\title{Probing lepton flavor violating dark matter scenarios via astrophysical photons and positrons}

\author{Jin-Han Liang\,\orcidlink{0000-0002-6141-216X}}
\email{jinhanliang@m.scnu.edu.cn}
\author{Yi Liao\,\orcidlink{0000-0002-1009-5483}}
\email{liaoy@m.scnu.edu.cn}
\author{Xiao-Dong Ma\,\orcidlink{0000-0001-7207-7793}}
\email{maxid@scnu.edu.cn}
\affiliation{State Key Laboratory of Nuclear Physics and
Technology, Institute of Quantum Matter, South China Normal
University, Guangzhou 510006, China}
\affiliation{Guangdong Basic Research Center of Excellence for
Structure and Fundamental Interactions of Matter, Guangdong
Provincial Key Laboratory of Nuclear Science, Guangzhou
510006, China}

\begin{abstract}
In this Letter we explore, for the first time, the constraints on lepton flavor violating (LFV) dark matter (DM) scenarios via the astrophysical photons and positrons, including both the annihilation and decay modes,
${\tt DM(+DM)}\to e^\pm \mu^\mp, e^\pm \tau^\mp, \mu^\pm \tau^\mp$.
Given the presence of LFV interactions in various DM models and the challenge of probing such interactions at terrestrial facilities, such as DM direct detection and collider experiments, indirect detection offers a unique approach to investigating them. We utilize the currently available photon datasets from the XMM-Newton, INTEGRAL, and Fermi-LAT telescopes, along with the positron datasets from the AMS-02 satellite, to establish stringent bounds on the relevant annihilation cross sections or decay widths. In particular, we include contributions to the photon spectrum from final state radiation, radiative decays, and inverse Compton scattering. We find that the INTEGRAL (AMS-02) provides the most stringent bound on the annihilation cross sections and decay widths for DM mass below (above) approximately 20 GeV, which are comparable to those of their lepton flavor conserving counterparts. 
\end{abstract}

\maketitle 

\vspace{0.1cm}{\bf Introduction.}
The nature of dark matter (DM) remains one of the most profound mysteries in modern physics. Despite overwhelming gravitational evidence for its existence, the particle properties of DM are still unknown \cite{Young:2016ala,Arbey:2021gdg}. 
Various DM candidates have been proposed 
over the past decades, each capable of inducing different types of interactions with the standard model (SM) particles that can be tested in diverse environments \cite{Bertone:2004pz,Feng:2010gw}. 
In general, DM candidates can be either stable due to an underlying symmetry or unstable but with a lifetime much longer than the age of the Universe, allowing for interactions with the SM particles  through either a pair of DM fields or a single DM field. Due to our limited knowledge of DM properties, a viable and compelling choice is to explore all possible interactions thoroughly. 

The DM particles may interact with the SM leptons, quarks, gauge, and Higgs bosons.
Over the years, DM direct and indirect detection experiments, along with collider searches, have extensively explored lepton- and quark-flavor-conserving DM-SM interactions
\cite{Roszkowski:2017nbc,Arcadi:2017kky,Boveia:2018yeb,Schumann:2019eaa,Cirelli:2024ssz,Liang:2024ecw,Nguyen:2024kwy,Liang:2025kkl}. Although no positive signals have been found, stringent bounds have been placed on the relevant observables and couplings. 
More recently, flavor anomalies observed in processes such as the meson decays $B^+\to K^+\nu\bar\nu$ from Belle II measurement \cite{Belle-II:2023esi} and 
$K^+\to \pi^+\nu\bar\nu$ by NA62 \cite{NA62:2024pjp} have stimulated interest in flavor-violating DM-quark interactions. This is because light DM particles can mimic neutrino pairs, and thus potentially enhance the decay rates to the experimentally required levels \cite{He:2023bnk}. 

However, lepton-flavor-violating (LFV) interactions between DM particles and the SM charged leptons have remained largely unexplored to date.
These interactions are particularly intriguing due to their distinctive experimental signatures and potential deep connection to the underlying mechanism responsible for the tiny neutrino mass. 
Many new physics scenarios that simultaneously explain neutrino mass and DM, such as the well-known scotogenic neutrino mass model \cite{Tao:1996vb,Ma:2006km}, can naturally induce such LFV DM interactions. Furthermore, LFV interactions may also play a role in the production of DM in the early Universe~\cite{Acaroglu:2022hrm,Acaroglu:2022boc,Acaroglu:2023cza}.

Experimentally, LFV DM scenarios are challenging to test in direct detection and collider experiments. 
In direct detection, DM particles near the Earth do not possess sufficient kinetic energy to overcome the mass gap between electrons and muons or taus, rendering LFV transitions inaccessible. 
In the absence of an electron-muon collider and alike, LFV DM interactions may also induce a process such as $e^-e^+\to\mu^\pm +e^\mp+\textrm{DM}+\textrm{DM}$ in an electron collider, which however is suppressed by additional SM couplings and phase space factors.
Although such interactions are difficult to investigate through terrestrial experiments, DM indirect detection offers a unique opportunity to explore them. In particular, DM particles that annihilate or decay into pairs of charged leptons with different flavors can produce characteristic astrophysical photon or positron signals. 
These secondary particles can be detected by space-based telescopes or detectors, thus providing a promising approach to place meaningful bounds on these LFV scenarios.  
On the other hand, if such flavored DM is sufficiently light with sub-GeV masses, the charged LFV decay $\ell_i\to \ell_j + {\tt DM+DM}$ can also impose constraints on the relevant interactions \cite{Jahedi:2025hnu}.

In this Letter, we fill the gap in probing the LFV DM interactions via indirect detection.
For both annihilating and decaying DM scenarios we derive model-independent constraints on all three final-state configurations:  
${\tt DM(+DM)}\to e^\pm\mu^\mp,~e^\pm \tau^\mp,~\mu^\pm \tau^\mp$.
We perform a comprehensive analysis of the resulted photon and positron spectra.  
By utilizing the available diffuse $X$/gamma-ray data from the XMM-Newton, INTEGRAL, and Fermi gamma-ray telescopes, along with positron measurements from AMS-02, we derive stringent bounds on the annihilation cross sections and decay widths. Our results provide the first systematic constraints on LFV DM interactions from astrophysical observations, thereby complementing existing limits on lepton-flavor-conserving (LFC) DM-SM interactions. To encourage further research toward this direction, we also present a UV-complete model to realize these scenarios.

\vspace{0.1cm}{\bf The prompt photon flux.}
There are two main contributions to the diffuse $X$/gamma-ray background originating from DM annihilation or decay in the late Universe: the localized galactic contribution and the extragalactic contribution due to the smooth distribution of the DM throughout the Universe. 
For the telescopes of interest, the dominant contribution comes from the Milky Way DM halo, 
while the extragalactic component is negligible.  
The photon flux due to the LFV DM scenario under consideration consists of three main components of photons from final state radiation ({\tt FSR}), radiative decay ({\tt Rad}), and inverse Compton scattering ({\tt ICS}). 
The total differential photon flux is 
\begin{align}
{d^2\Phi_\gamma \over d\Omega dE_\gamma}  
= {d^2\Phi_\gamma^{\tt FSR} \over d\Omega dE_\gamma} 
+ {d^2\Phi_\gamma^{\tt Rad} \over d\Omega dE_\gamma} 
+ {d^2\Phi_\gamma^{\tt ICS} \over d\Omega dE_\gamma}.
\end{align}
In this work, we focus on the dominant s-wave annihilation or decay scenarios, in which the particle physics details and the DM astrophysical distributions can be factorized. 
For the prompt photon component from {\tt FSR} and {\tt Rad}, the flux takes the following form,
\begin{align}
{d^2\Phi_\gamma^I \over d\Omega dE_\gamma} =
\frac{1}{4\pi}
\frac{d N_\gamma^I}{dE_\gamma} 
\begin{cases}
 \displaystyle \frac{ \langle \sigma v \rangle_{ij}}{2 m_{\texttt{DM}}^2}
\int_{\rm l.o.s}\rho^2(\pmb{r}) d s, & ({\tt A}), 
\\[7pt]
\displaystyle \frac{\Gamma_{ij}}{m_{\texttt{DM}} }  
\int_{\rm l.o.s}\rho(\pmb{r})  d s,  &({\tt D}), 
\label{eq:dPhidEprompt}
\end{cases}
\end{align}
where $d N_\gamma^I/dE$ denotes the normalized photon spectrum in each annihilation or decay process, with $I$ representing either {\tt FSR} or {\tt Rad}. 
$\langle \sigma v \rangle$ denotes the velocity-averaged annihilation ({\tt A}) cross section for ${\tt DM+DM}\to \ell_i^- \ell_j^+$, 
while $\Gamma$ represents the decay width for ${\tt DM}\to \ell_i^- \ell_j^+$ in the decaying DM scenario ({\tt D}). 
Note that the above annihilation formula applies to real scalar/vector and Majorana fermion DM cases, whereas a factor of 1/2 has to be included for complex scalar/vector and Dirac fermion DM cases. 
$\rho(\pmb{r})$ is the DM energy density and the standard Navarro-Frenk-White profile is adopted \cite{Navarro:1995iw,Navarro:1996gj}. The integration of $s$ is along the line of sight (l.o.s) that accounts for all the contributions along a specific direction. 

Now we turn to the determination of the photon spectrum. 
For the {\tt FSR}, the photon spectrum from s-wave annihilation or decay shows only mild dependence on the interaction structure between DM and leptons, particularly when the DM mass is well above the threshold of the final-state particles \cite{Coogan:2019qpu}. Therefore, for concreteness, we adopt the normalized {\tt FSR} spectrum due to the local operator $\bar{\ell_i} \gamma_\mu \ell_j \bar{\chi} \gamma^\mu\gamma_5 \chi$ for a Majorana DM $\chi$ in our analysis. 
The details of our calculation for the normalized photon spectra and their weak dependence on the operator structure will be presented 
in our forthcoming long paper~\cite{Jahedi:2026dzy}.

The {\tt Rad} photons originate from the radiative leptonic and/or hadronic decays of the initially produced charged leptons from DM annihilation or decay. The {\tt Rad} photon spectrum from the four-body radiative decay process $\ell^- \to \ell'^{-} \bar\nu_{\ell'} \nu_\ell \gamma$ (or its charge-conjugation counterpart) is given by~\cite{Essig:2009jx,Kuno:1999jp},
\begin{align}
{d N_\gamma^{{\tt Rad},\ell} \over d E_\gamma } 
& = 
{\alpha (1-x_\ell)\over 36\pi E_\gamma}
\Big\{
x_\ell(1-x_\ell)(46-55x_\ell)-102
\nonumber
\\
& + 12\big[3-2x_\ell (1-x_\ell)^2\big]\ln{{1-x_\ell\over r_{\ell'\ell}}}
\Big\},
\end{align}
where $x_\ell = 2E_\gamma/ m_\ell$ and $r_{\ell'\ell}= (m_{\ell'}/m_\ell)^2$ with $\ell\ell'=(\mu e,\tau e,\tau \mu)$.
In addition, the hadronic decays of the $\tau$ lepton can also contribute significantly to its photon spectrum, and we obtain the relevant spectrum by \textsc{Pythia} simulations \cite{Bierlich:2022pfr}. 
For the produced charged leptons $\ell_i^-$ and $\ell_j^+$ in flight, the above spectrum should be boosted to the DM rest frame in which
$E_{i,j}^{\tt A} = m_{\tt DM} \pm (m_i^2-m_j^2)/(4m_{\tt DM})$  
or $E_{i,j}^{\tt D} = m_{\tt DM}/2 \pm (m_i^2-m_j^2)/(2m_{\tt DM})$. 

\vspace{0.1cm}{\bf The {\tt ICS} photon flux.}
The {\tt ICS} photons are from scattering of initial high energy electrons ($e^-$) or positrons ($e^+$) off the galactic background photons mainly consisting of the cosmic microwave background (CMB), rescattered infrared light by dust, and optical starlight. 
Its flux can be written as 
\begin{align}
{d^2\Phi_\gamma^{\tt ICS} \over d\Omega dE_\gamma} = {1\over 4\pi}\int_{\tt l.o.s} ds     
\frac{j(E_\gamma, \pmb{r})}{E_\gamma}, 
\end{align}
where $j(E_\gamma, \pmb{r})$ is the photon emissivity at position $\bm{r}$, obtained by convolving the $e^\pm$ spectral number density ($dn_{e^\pm}/dE_e$) with the corresponding radiation power into photons
[${\cal P}_{\tt ICS}(E_\gamma, E_e, \bm{r})$],
\begin{align}
j(E_\gamma, \pmb{r}) = \int_{m_e}^{E_e^{\tt max}} d E_e   
~{\cal P}_{\tt ICS}(E_\gamma, E_e, \bm{r}) 
\frac{dn_{e^\pm}}{dE_e}(E_e, \pmb{r}),
\end{align}
with the maximum energy $E_e^{\tt max} = {\rm max}\big[ E_i^{\tt A(D)}, E_j^{\tt A(D)}\big]$ for the annihilation (decay) scenario. Here $dn_{e^\pm}/dE_e$ includes contributions from both $e^-$s and $e^+$s produced via the annihilation (decay) mode ${\tt DM(+DM)}\to \ell_i^- \ell_j^+$, and
\begin{align}
{\cal P}_{\tt ICS} = \int d \epsilon_\gamma  
(E_\gamma-\epsilon_\gamma)\frac{dn_\gamma}{d\epsilon_\gamma}(\epsilon_\gamma, \bm{r}) \frac{d\sigma_{\rm eff}}{d E_\gamma}(\epsilon_\gamma,E_\gamma, E_e).    
\end{align}
Here, $\epsilon_\gamma$ denotes the photon energy before the scattering, 
and the background photon number spectral distribution $dn_\gamma/d\epsilon_\gamma$ includes the three components mentioned above. The CMB component is given by the thermal Bose-Einstein distribution at $T=2.73\,\rm K$. 
For the infrared and starlight components, we use the interstellar radiation field maps extracted from \cite{Vladimirov:2010aq}. 
$d\sigma_{\rm eff}/d E_\gamma$ is an effective ICS cross section that accounts for boost and reference transformation effects, and its expression can be found in Eq.\,(2.48) of \cite{Blumenthal:1970gc}.

For the $e^\pm$ spectral distribution, we adopt the ``on the spot'' approximation, i.e., neglecting the electron/positron diffusion effects and assuming that the scattering occurs at the same location.
The diffusion can shift the photon flux from the galactic center region outward, with the flux from fully numerical \textsc{Galprop} calculations~\cite{Cholis:2008wq,Borriello:2009fa} agreeing with the approximation within a factor of two~\cite{Meade:2009iu,Cirelli:2009vg}. These $e^\pm$ produced from DM annihilation or decay take the form~\cite{Cirelli:2009vg,Cirelli:2020bpc},
\begin{align}
\frac{dn_{e^\pm}}{dE_e}(E_e, \pmb{r}) = 
\frac{Y_{e^\pm}(E_e)}{b_{\tt tot}(E_e,\bm{r})}
\begin{cases}
 \displaystyle \frac{ \langle \sigma v \rangle_{ij} }{2 m_{\texttt{DM}}^2}\rho^2(\pmb{r}), & ({\tt A}), 
\\[7pt]
\displaystyle  \frac{\Gamma_{ij}}{m_{\texttt{DM}} }  
\rho(\pmb{r}),  &({\tt D}), 
\label{eq:dPhidEprompt}
\end{cases}
\end{align}
where $b_{\rm tot}(E_e, \pmb{r})$ is the total energy loss function of the $e^\pm$s \cite{Blumenthal:1970gc}.
The yield of $e^\pm$ is given by
\begin{align}
Y_{e^\pm}(E_e) = \int_{E_e}^{E_e^{\tt max}} d \tilde{E}_e \frac{dN_{e^\pm}}{d\tilde{E}_e}(\tilde{E}_e).
\end{align}
Here, $dN_{e^\pm} / d\tilde{E}_e$ are the energy spectra of electrons and positrons produced per DM annihilation or decay event,
including contributions from the primary $e^\pm$ (when $\ell_i^-=e^-$ or $\ell_j^+=e^+$) and the secondary ones due to decays of $\mu$ and $\tau$.
The spectrum of primary $e^\pm$ is simply represented by a Dirac delta function.
For muon decay at rest, the $e^\pm$ spectrum is determined by the process $\mu^- \to e^- \nu_\mu \bar{\nu}_e$ and its conjugate.
For $\tau$ decay at rest, we obtain the $e^\pm$ spectrum by using \textsc{Pythia8} simulations.
A comparison of photon components contributing to the predicted photon fluxes from the three processes will be detailed in~\cite{Jahedi:2026dzy}.

\vspace{0.1cm}{\bf Positron flux.}
The positron flux resulting from DM annihilation or decay is given by \cite{Cirelli:2010xx}:
\begin{align}
\frac{d^2 \Phi_{e^+}}{d\Omega dE_e } = 
\frac{1}{4\pi\, b_{\tt tot}}
   \int_{E_e}^{E_e^{\tt max}} d \tilde{E}_e  Q_{e} (\tilde{E}_e, \pmb{r}) I(E_e, \tilde{E}_e, \pmb{r}),
\end{align}
where the source term $Q_{e}$ is defined as
\begin{align}
Q_e(\tilde{E}_e, \pmb{r}) = \frac{dN_{e^+}}{d \tilde{E}_e} \times
\begin{cases}
\displaystyle \frac{\langle \sigma v \rangle_{ij} }{2 m_{\texttt{DM}}^2}\rho^2(\pmb{r}), & ({\tt A}), 
\\[7pt]
\displaystyle \frac{\Gamma_{ij} }{m_{\texttt{DM}}}  \rho(\pmb{r}), &  ({\tt D}).
\end{cases}
\end{align}
The generalized dimensionless halo function $I(E_e, \tilde{E}_e, \pmb{r})$ serves as a Green's function, 
describing the probability that a positron injected with energy $\tilde{E}_e$ is observed with $E_e$ at position $\pmb{r}$.
We evaluate the expression at $\pmb{r} = \pmb{r}_\odot$ with $\pmb{r}_\odot$ being the location of the Sun.

\vspace{0.1cm}{\bf Datasets.}
We consider the three $X$/gamma-ray telescopes---INTEGRAL, XMM-Newton, and Fermi-LAT, along with the AMS-02 experiment to establish our constraints. Below, we provide a brief description of their datasets used in our analysis. 

{\it INTEGRAL}. We use observational data from the INTEGRAL/SPI $X$-ray spectrometer reported in Ref.~\cite{Bouchet:2011fn}. The dataset spans the period from 2003 to 2009, corresponding to a total exposure time of approximately $10^{8}\,\rm s$. 
Specifically, we extract the flux measurements presented in Fig.\,5 of Ref.\,\cite{Bouchet:2011fn}.
The data are divided into five energy intervals: 27–49 keV, 49–90 keV, 100–200 keV, 200–600 keV, and 600–1800 keV. For the first four intervals, the sky region is partitioned into 21 bins along the galactic latitude, with the flux in each bin integrated over the longitude range of 
$|l| < 23.1^\circ$. For the highest energy interval, 15 latitude bins are used, and the fluxes are integrated over a wider longitude range of 
$ |l| < 60^\circ$.
In our analysis, the three central latitude bins are removed due to significant background contamination and substantial uncertainties in the predicted photon fluxes \cite{Cirelli:2020bpc}.

{\it XMM-Newton}. All-sky observations were performed using the MOS and PN cameras onboard the \textit{XMM-Newton} satellite, covering an energy range of approximately 0.2–20\,keV.
The data with point sources removed are publicly available in~\cite{Foster:2021ngm,XMM_BSO_DATA}, for which the sky is divided into 30 concentric rings defined by the angle $\psi$, excluding the latitude region $|b| \leq 2^\circ$.
For each ring, the exposure time, observed event counts, instrumental response matrices, and energy bins before and after detector effects (input and output bins, respectively) are provided in~\cite{XMM_BSO_DATA}. 
Please note that the data are normalized to the exposure-weighted solid angle, which accounts for the telescope’s nonuniform exposure across the ring due to its observational footprint, rather than the instrument’s geometric solid angle. To properly interpret the data, one can refer to the {\it python} notebook in \cite{XMM_BSO_DATA}. 
Without accounting for this normalization, the resulting constraints would appear artificially stronger by several orders of magnitude as pointed out in~\cite{Balaji:2025afr}.

The number of photon events in the output $i$th bin is
\begin{align}
N_\gamma^i = \text{exposure} \times \sum_j R_{ij} \, \phi_\gamma^j,
\end{align}
where $\phi_\gamma^j$ is the integrated DM-induced photon flux in the input $j$th bin in units of $\mathrm{cm^{-2}s^{-1}}$, and $R_{ij}$ is the detector response matrix element (in units of cm$^2$) with the effective area already included.
To avoid the dominant instrumental background, the energy range of the output bins is restricted to 2.5--8~keV for the MOS camera and 2.5--7~keV for the PN camera~\cite{Foster:2021ngm}.

{\it Fermi-LAT}. We use Fermi 2012 datasets covering $0 < l < 360^\circ$ and $8^\circ < |b| < 90^\circ$, in the energy range from 200 MeV to 10 GeV, as shown in the upper panel of Fig.\,(12) in~\cite{Fermi-LAT:2012edv}, to derive our bounds.
The latitude range is chosen to enhance the signal-to-background ratio while minimizing the uncertainty in the DM profile \cite{Essig:2013goa}.

{\it AMS-02}. Precision measurements of the cosmic-ray positron spectrum up to 1\,TeV have been conducted by AMS-02 onboard the International Space Station, based on 1.9 million events collected from May 19, 2011 to November 12, 2017 \cite{AMS:2019rhg}. In this work, we utilize the data presented in Table SI of the supplementary material in 
\cite{AMS:2019rhg}. To reduce the influence of solar modulation, which introduces sizable uncertainties at low energies, only data with energies above 20\,GeV are used in our analysis.
The datasets we extracted from INTEGRAL, Fermi-LAT, and AMS-02 can be found in the ancillary files on \href{https://arxiv.org/abs/2508.05121}{arXiv}.

\vspace{0.1cm}{\bf Statistics.}
Due to significant uncertainties in astrophysical processes and observations, we adopt a conservative $\chi^2$ statistic method to enhance the reliability of our results \cite{Cirelli:2020bpc,Cirelli:2023tnx},
\begin{align}
\chi^2 \equiv \sum_i \Big( \frac{{\tt max}\left[ S_i (w, m_{\rm DM}) - O_i, 0 \right]}{\sigma_i} \Big),
\label{eq:sta}
\end{align}
where $S_i$ and $O_i$ are the DM-induced and observed photon fluxes or event counts in the $i$th bin with uncertainty $\sigma_i$, and $w = \langle \sigma v \rangle\,(\Gamma)$ stands for the annihilating (decaying) DM scenario. 
In this work, we adopt a background-free assumption and take all observed fluxes as originating from DM annihilation to derive a conservative limit.
For each DM mass point, we derive the $2\sigma$ bound on the parameter $w$ by requiring $\chi^2 = 4$.

\begin{figure}[t]
\centering
\includegraphics[width=0.4\textwidth]{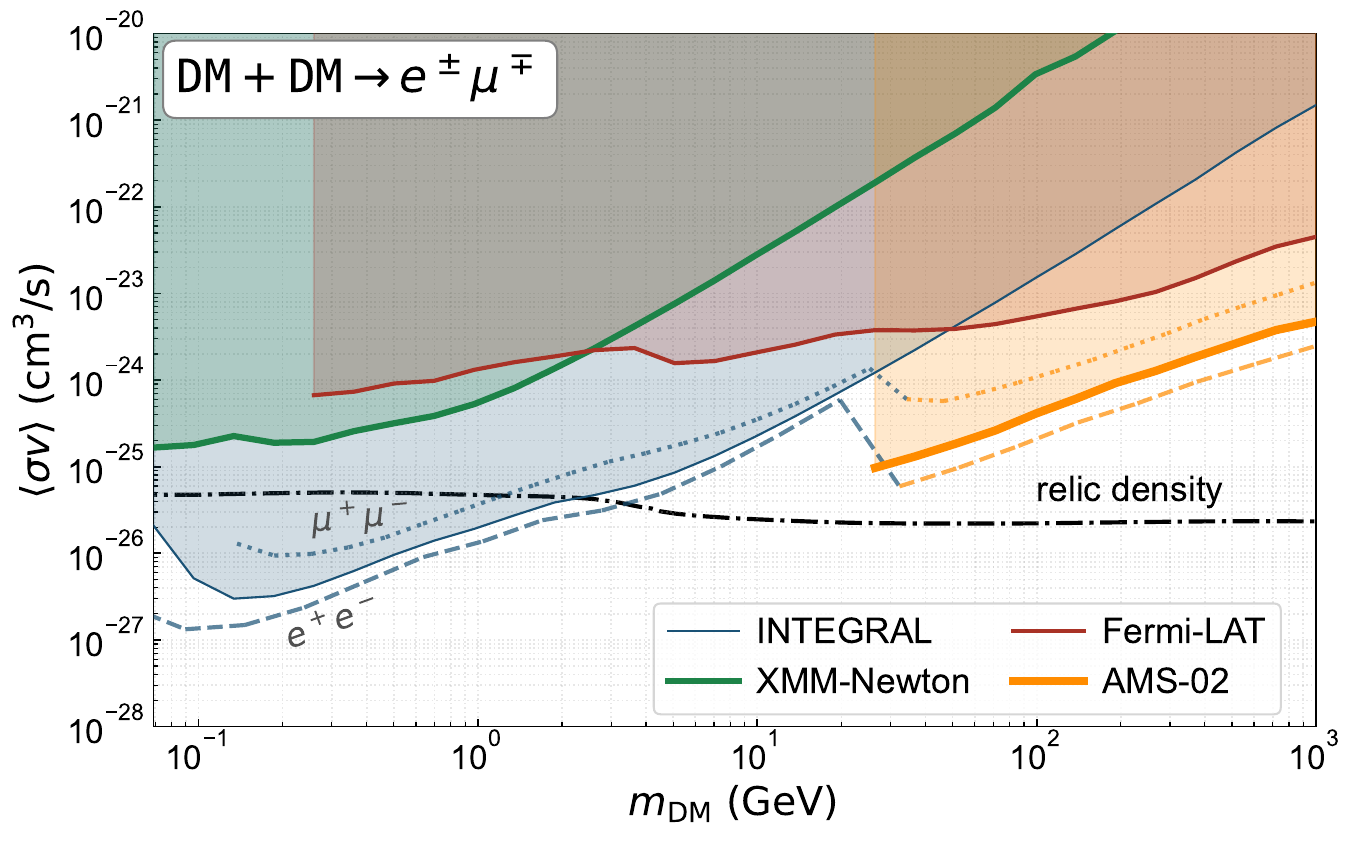}
\\
\includegraphics[width=0.4\textwidth]{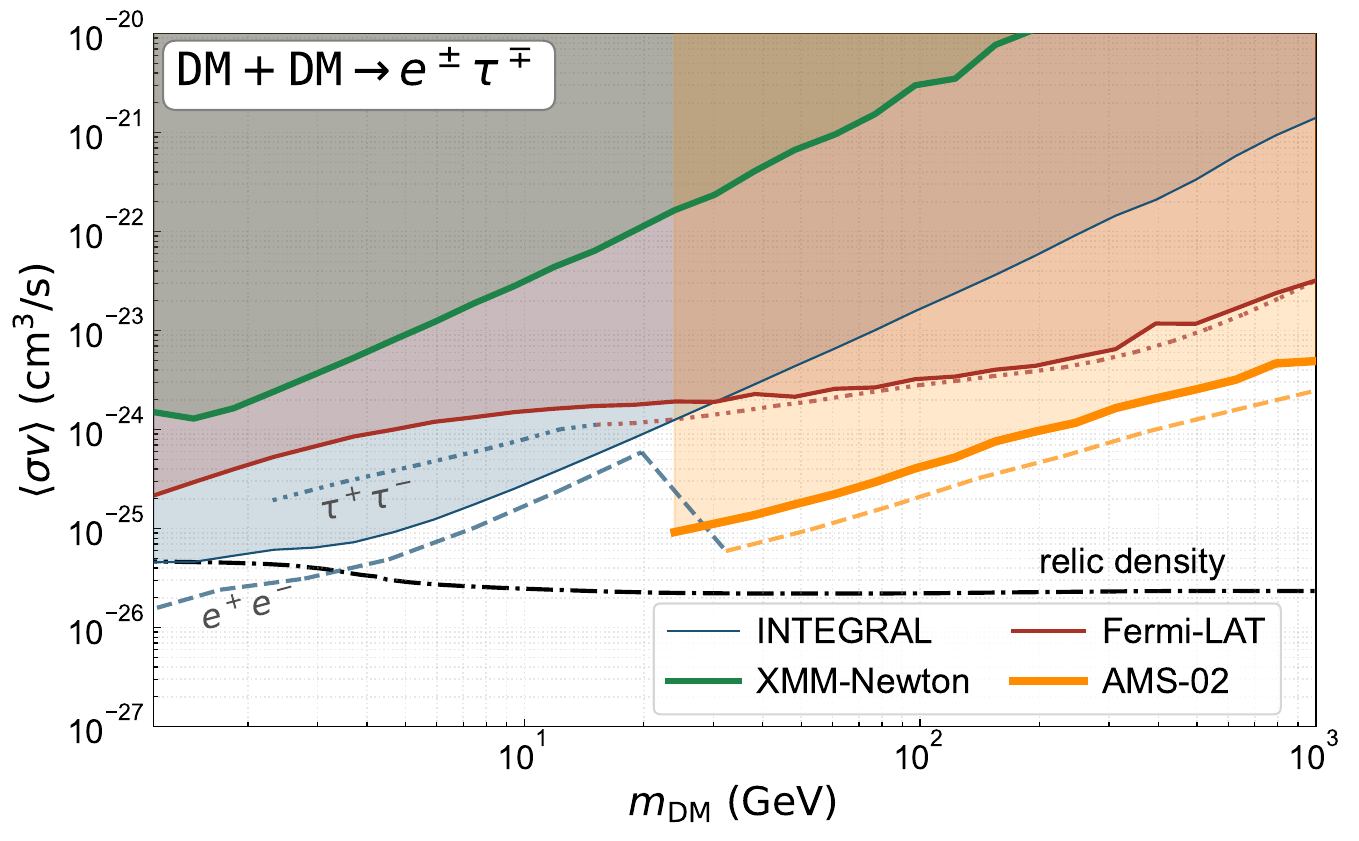}
\\
\includegraphics[width=0.4\textwidth]{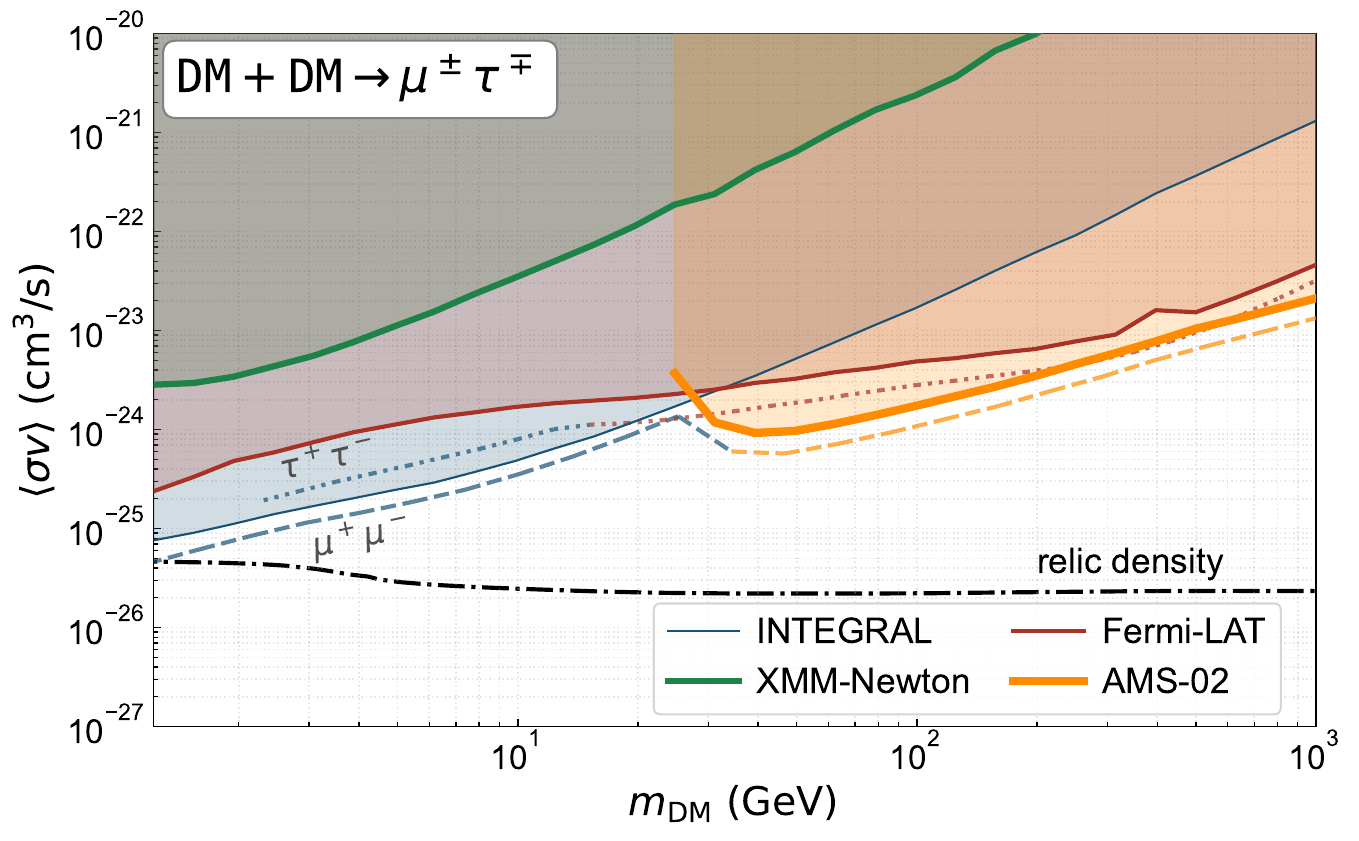}
\vspace{-0.2cm}
\caption{
Constraints on DM LFV annihilation cross section $\langle\sigma v\rangle_{ij+ji}$ from INTEGRAL (blue), XMM-Newton (green), Fermi-LAT (red), and AMS-02 (orange). For each LFV $\ell_i^\pm \ell_j^\mp$ final state, we also present the combined most stringent constraints on the LFC cases $\ell_i^+ \ell_i^-$ and $\ell_j^+ \ell_j^-$, where the corresponding color 
in each mass region indicates the experiment responsible for the constraint.
The black dot-dashed curves show the corresponding thermal average of $\sigma v$ required to generate the observed DM relic density~\cite{Steigman:2012nb}.
}
\label{fig:ann}
\end{figure}

\begin{figure}[t]
\centering
\includegraphics[width=0.4\textwidth]{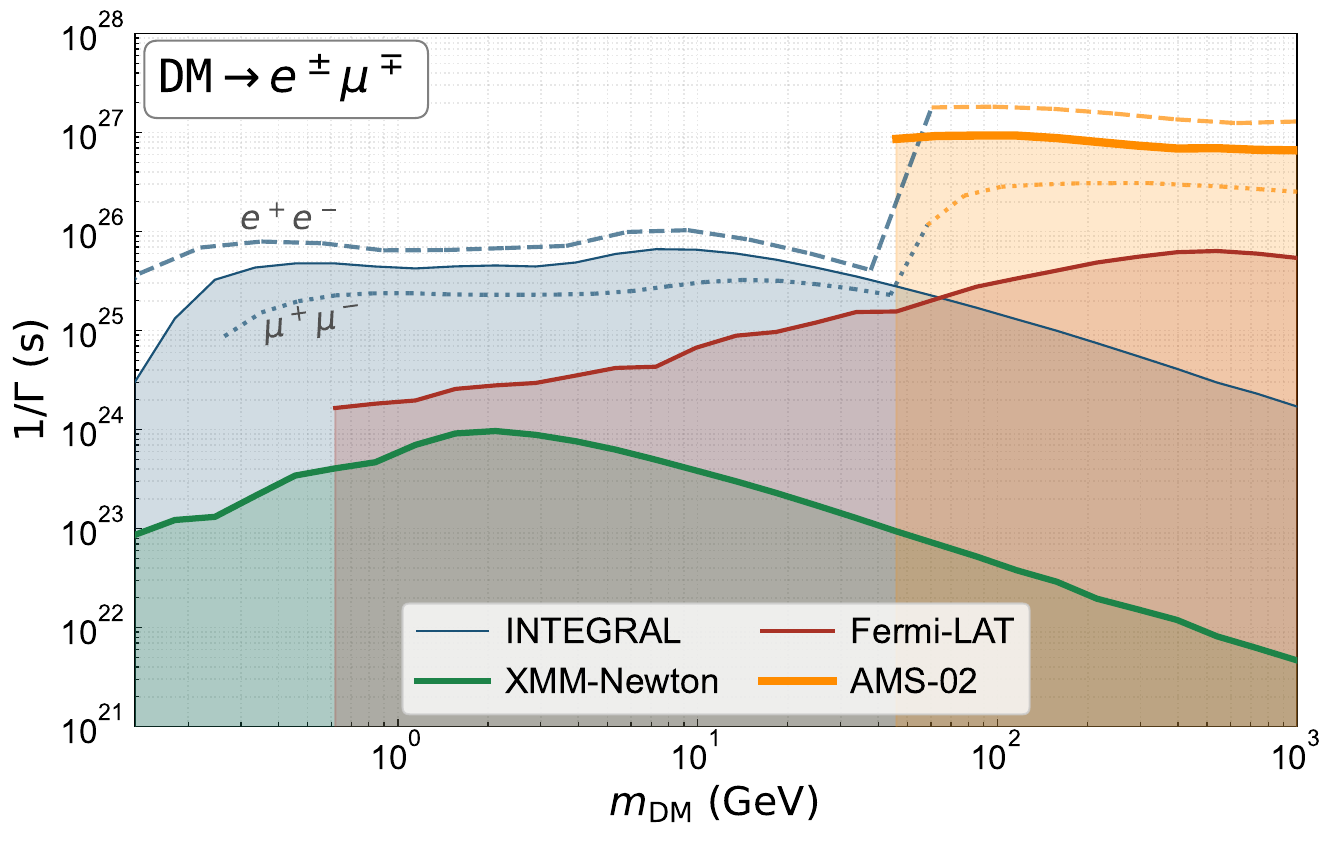}
\\
\includegraphics[width=0.4\textwidth]{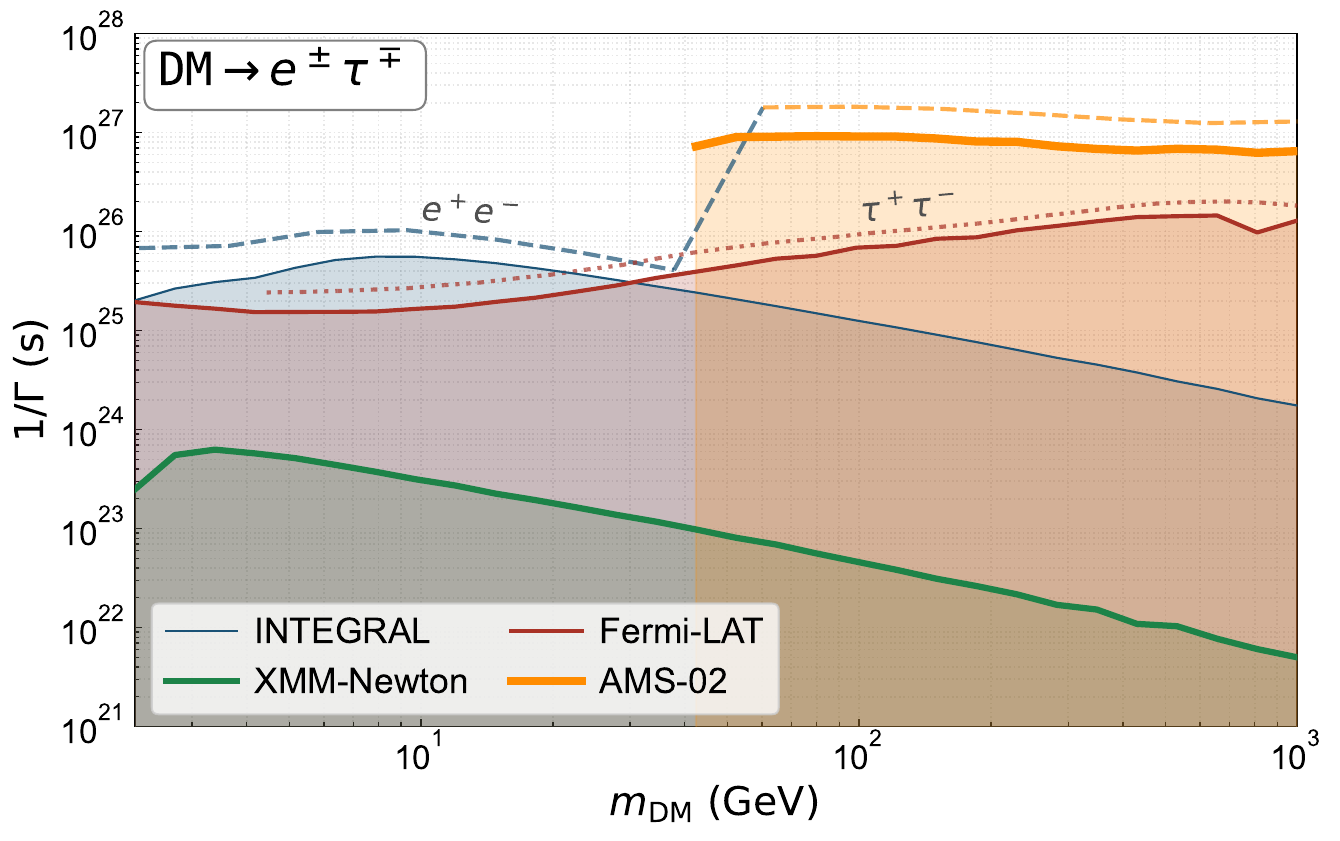}
\\
\includegraphics[width=0.4\textwidth]{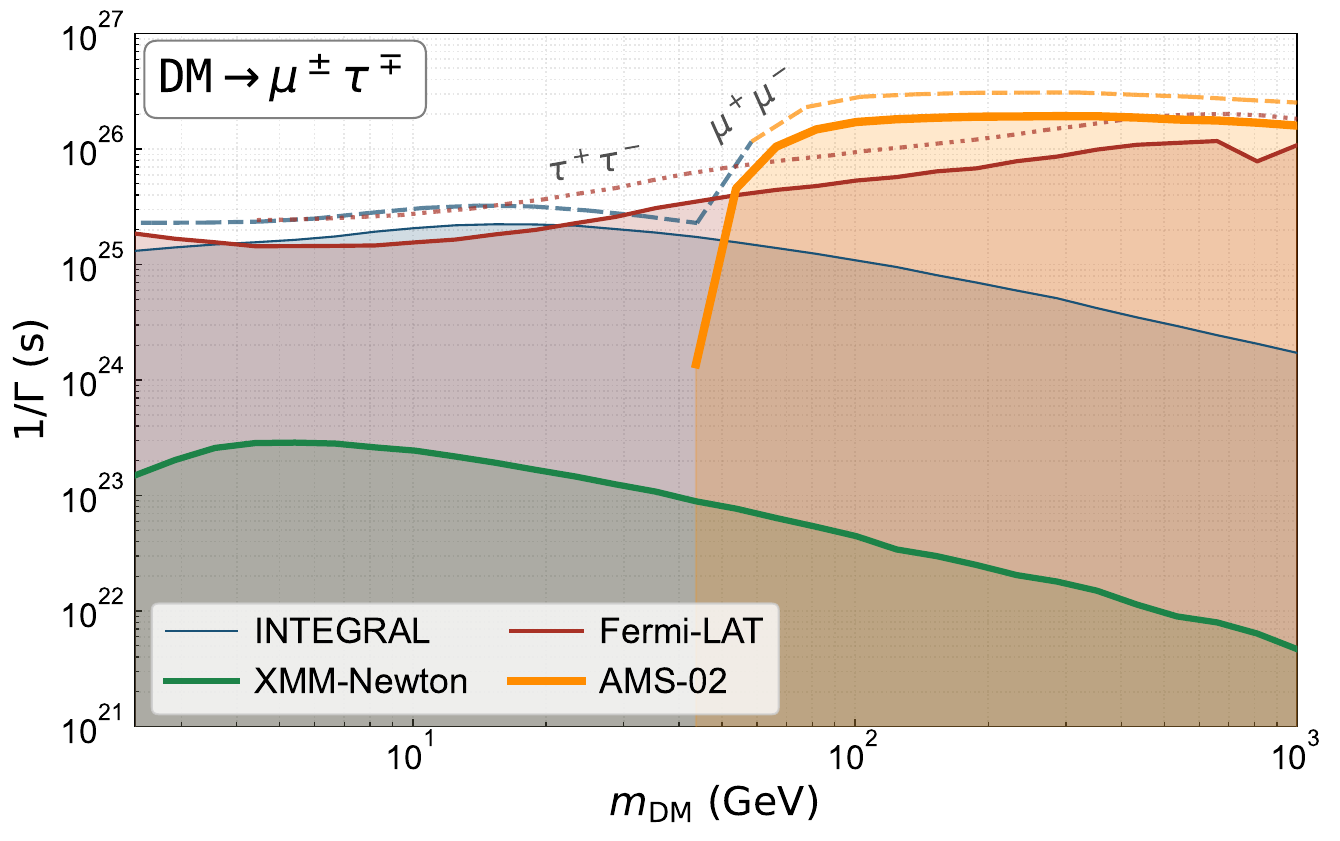}
\vspace{-0.2cm}
\caption{
Similar to \cref{fig:ann}, but for the constraints on the inverse decay width $1/\Gamma_{ij+ji}$ in LFV decaying DM scenarios.}
\label{fig:dec}
\end{figure}

\vspace{0.1cm}{\bf Results.}
Based on the theoretical formalism and datasets discussed above, 
our final constraints are shown in \cref{fig:ann} for the annihilating DM scenario and in \cref{fig:dec} for the decaying DM scenario, respectively. 
Besides the LFV cases, we have also calculated the constraints for the LFC cases from each experiment to compare with the results in the literature and validate our approach. 
Our constraints on the $e^+ e^-$ and $\mu^+ \mu^-$ channels from INTEGRAL and Fermi-LAT datasets are consistent with those reported in \cite{Cirelli:2020bpc,Cirelli:2023tnx,Essig:2013goa}.
Regarding the XMM-Newton constraints, our results are approximately three orders of magnitude weaker than those in \cite{Cirelli:2023tnx}, due to our proper inclusion of the geometric factor discussed in \cite{Balaji:2025afr}.

Since the photon flux and the $e^\pm$ yield from the LFV $\ell_i^\pm \ell_j^\mp$ final state fall between those of the LFC $\ell_i^+ \ell_i^-$ and $\ell_j^+ \ell_j^-$ final states,
the constraints on each LFV channel from a given experiment generally lie between the constraints on the corresponding two LFC channels. These characteristics are clearly illustrated in \cref{fig:ann,fig:dec} and further strengthen the robustness of our results.

{\it Annihilating DM scenario}. As shown in \cref{fig:ann},
the most stringent constraints in each flavor combination are set by the INTEGRAL data for $m_{\tt DM} \lesssim {\cal O}(10)~\rm GeV$, and by the AMS-02 data for $m_{\tt DM} \gtrsim {\cal O}(10)~\rm GeV$. 
Moreover, the INTEGRAL constraints are dominated by the ICS photons.
AMS-02 measures final-state positrons, making the $\mu\tau$ channel less constrained at low DM masses than the $e\mu$ and $e\tau$ channels, where positrons are produced directly. In contrast, $\mu\tau$ positrons come only from secondary decays, yielding lower energies.

{\it Decaying DM scenario}.
For the $\mu^\pm \tau^\mp$ channel, it is notable from \cref{fig:dec} that when $m_{\tt DM} \lesssim 30~\mathrm{GeV}$, the resulting constraints are weaker than those from either the $\mu^+ \mu^-$ or $\tau^+ \tau^-$ channels.
This behavior arises because the INTEGRAL and Fermi-LAT constraints are comparable in strength in this mass range.
Moreover, the $e^\pm$ yield from $\mu^\pm \tau^\mp$ is smaller than that from $\mu^+ \mu^-$, while the associated photon flux is lower than that from $\tau^+ \tau^-$, leading to overall weaker indirect detection sensitivity.

\vspace{0.1cm}{\bf A simple UV model.}
We provide a simple model to realize the LFV annihilating DM scenario by extending the SM with an additional Higgs doublet $\Phi$ and a scalar DM candidate $\phi$. We work in the so-called Higgs basis in which $\Phi$ does not develop a vacuum expectation value (VEV).
We assume that $\Phi$ couples predominantly to the SM lepton sector and impose a $\mathbb Z_2$ symmetry under which $\phi$ is odd while all other fields are even. 
Then the relevant Lagrangian terms are:
\begin{align}
 {\cal L}_{\Phi, \phi} \supset y_{ij}\overline{L_i}e_j \Phi - \lambda_{H\Phi\phi}  \Phi^\dagger H \phi^2 +\rm H.c.,
 \label{eq:UVLag}
\end{align}
where $L$ and $e$ are the SM left- and right-handed leptons, and $H$ is the SM Higgs doublet with VEV $v$. 

By assuming dominance of off-diagonal over diagonal elements of the Yukawa couplings $y_{ij}$, the DM were produced via the LFV annihilation modes in the early universe through the thermal freeze-out. The annihilation cross section for the process $\phi\phi\to \ell_i^-\ell_j^+$ ($i\neq j$) is
\begin{align}
\sigma_{ij} 
& =  \frac{1}{16\pi s }
\frac{\lambda^{1/2}(s,m_i^2, m_j^2)}{\lambda^{1/2}(s, m_\phi^2, m_\phi^2)}
\frac{v^2 |\lambda_{H\Phi\phi}|^2}{(s-m_{\Phi}^2)^2+m_{\Phi}^2 \Gamma_{\Phi}^2 }
 \times
\\
&[(s- m_i^2 -m_j^2)
 (|y_{ij}|^2 +|y_{ji}|^2)
- 4 m_i m_j \Re(y_{ij} y_{ji})].
\nonumber
\end{align}
Here $\sqrt{s}$ is the center-of-mass energy, $m_a$ the masses of various particles  $(a=\ell_i,\ell_j,\phi,\Phi)$, $\Gamma_\Phi$ the width of $\Phi$, and $\lambda(x,y,z)$ the usual triangle function.

\begin{figure}
\centering
\includegraphics[width=0.4\textwidth]{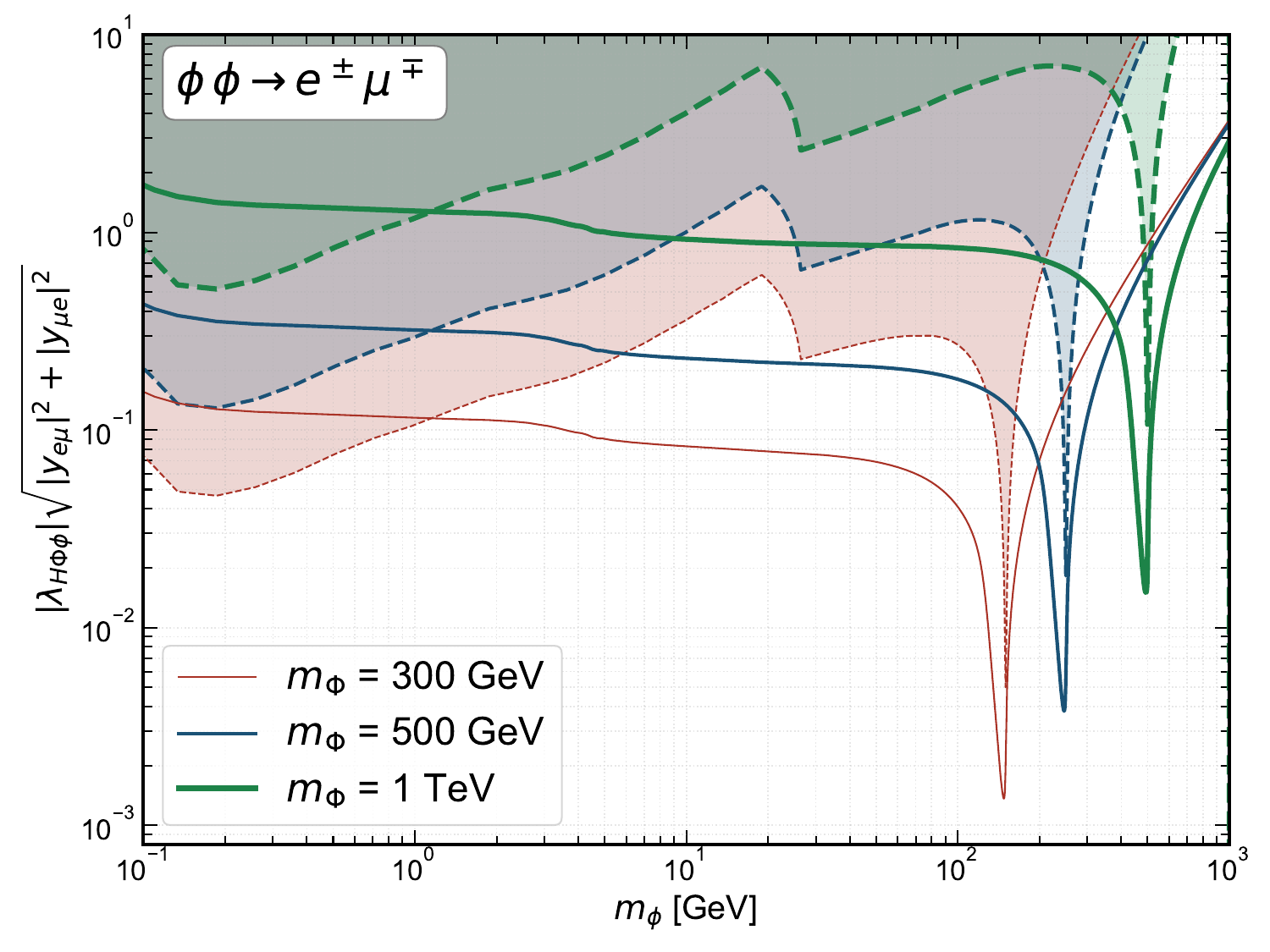}
\vspace{-0.2cm}
\caption{Parameter space of the relic density (solid curves) and the combined indirect detection constraints (dashed curves) for the DM annihilation channel 
$\phi \phi \to e^\pm \mu^\mp$.}
\label{fig:model}
\end{figure}

To illustrate the indirect detection constraints on the model parameters, we take the $e\mu$ flavor as an example. In \cref{fig:model}, the solid curves represent the coupling strength required to achieve the correct relic density as a function of the DM mass, while the colored regions are excluded by indirect detection constraints.
To obtain the relic density curves, we have used the velocity-averaged cross section formula developed in \cite{Gondolo:1990dk} and matched it to the required value calculated in \cite{Steigman:2012nb}. 
The dip corresponds to the point where $m_\phi\approx m_{\Phi}/2$, and is regulated by the estimated width $\Gamma_{\Phi}\sim 0.1 \,m_{\Phi}/(4\pi)$.
It can be seen that indirect detection constrains the DM mass to be above the GeV scale in this scenario. 
For the other two flavor scenarios ($e\tau$ and $\mu\tau$), the relic density curves remain applicable, but the lower bound of DM mass is approximately $m_\tau/2$ due to kinematic requirement. However, in these cases, the indirect constraints are weaker, indicating that the entire mass range is still viable.    
For the $e\mu$ case, our scenario can easily satisfy the constraints from the muonium oscillation, muon $g-2$, and forward-backward asymmetry of $e^+ e^- \to \mu^+ \mu^-$ derived in \cite{Endo:2020mev}.   
Finally, the LFV processes $\mu\to e\gamma$ or $\mu\to 3e$  involve new LFC couplings, which can be adjusted to sufficiently small values to comply with experimental constraints.

\vspace{0.1cm}{\bf Conclusion.}
In this Letter, we have thoroughly examined LFV DM scenarios through DM indirect detection. 
By exploiting the astrophysical $X$/gamma-ray data from XMM-Newton, INTEGRAL, Fermi-LAT, as well as the positron data from AMS-02, we have, for the first time, set stringent constraints on the DM annihilation cross section or decay rate for the three LFV channels, ${\tt DM(+DM)}\to e^\pm \mu^\mp, e^\pm \tau^\mp, \mu^\pm \tau^\mp$. 
Since such LFV DM scenarios have been overlooked in previous studies, we also present a UV-complete model to realize these interactions, demonstrating that such flavorful DM models remain viable to generate the observed relic abundance.

\section*{Acknowledgements}
We thank the anonymous referee for several comments and suggestions that have helped us improve the presentation of our work.
We thank Zihong Cheng for helpful discussions.
This work was supported by Grants  
No.\,NSFC-12305110, 
No.\,NSFC-12035008,
and No.\,NSFC-12347121.

\bibliography{paper_refs.bib}{}
\bibliographystyle{utphys}

\end{document}